\documentclass[10pt,conference]{IEEEtran}
\usepackage{graphicx}
\usepackage{ amsmath }
\usepackage{ amssymb }
\usepackage{amsfonts}
\newtheorem{thm}{Theorem}[section]
\newtheorem{prop}[thm]{Proposition}
\newtheorem{cor}[thm]{Corollary}

\newtheorem{example}[thm]{Example}

\newtheorem{rem}[thm]{Remark}
\begin{document}
\title{
On the non-existence for quantum LDPC codes of type IEEE802.16e with rates 1/2 and 2/3B}
\author{
\authorblockN{Manabu HAGIWARA}
\authorblockA{Research Center for Information Security (RCIS),\\
National Institute of\\
Advanced Industrial Science and Technology (AIST),\\
Akihabara-Daibiru 11F,\\
1-18-13 Sotokanda, Chiyoda-ku, Tokyo, Japan.\\
Email: hagiwara.hagiwara@aist.go.jp }
\and
\authorblockN{Hideki IMAI}
\authorblockA{Advanced Industrial Science and Technology (AIST)\\
and Chuo University,\\
1-13-27 Kasuga,
Bunkyo-ku,\\
Tokyo,
Japan.\\
Email: h-imai@aist.go.jp }
}
\maketitle

\section{Introduction}
In this paper, we discuss a construction of CSS codes derived from pairs of practical irregular LDPC codes.
Intersection studies between quantum error-correcting codes and LDPC codes are current hot research topics \cite{camara,mackay,entangle,qqcldpc,zhao,hagiwara,tillich,tan}.
One of aims of quantum error-correcting code theory is to construct quantum error-correcting codes with small length and high error-correcting performance.
The reason why LDPC codes are chosen is their high error-correcting performances as classical error-correcting codes.
In fact, LDPC codes have almost achieved a theoretical error-correction limit, called a Shannon limit, with various conditions, for example, code rate, code length, communicatin channel, and others.
This fact induces naturally derived the quantum error-correcting codes.

In the researches, we encounter a problems to hold a condition, which we call it the ``twisted condition'', for a pair of classical codes to derive a quantum code.
The condition requires a geometrically orderly structure to two classical code spaces.
More explicitly, the dual code of one of two codes must be contained in the other code.
On the other hand, a complicated structure, i.e. a random structure, is suitable for achieving high error-correcting performance.
The construction theory of classical codes for a CSS code is to satisfy conflicting, orderly and complicated, requirements.
We re-write the two conditions in terms of parity-check matrices.
The twisted condition implies the orthogonality of their parity-check matrices.
A complex structure implies that non-zero elements in the parity-check matrices should be irregularly arranged.

The previous researches on quantum LDPC codes have focused on quantum regular LDPC codes \cite{camara,mackay,hagiwara,tillich} as same as that of classical LDPC codes.
But now, a main stream of classical LDPC code researches is a construction method for irregular LDPC codes.
The crucial reason why the main research stream has changed is error-correcting performance.
In fact, the performances of irregular LDPC codes are extremely better than those of regular LDPC codes.
This fact is theoretically proved by density-evolution theory and the Gaussian approximation method \cite{richardson,chung}.

Our design of irregular LDPC codes is based the design written in the standardization of IEEE802.16e.
It is needless to say that irregular LDPC codes which are chosen in IEEE802.16e are designed well-considerably and show high error-correcting performance with practical length.
Our research has tried to make a CSS code with a pair of LDPC codes of type IEEE802.16e.
To our regret, we proved that it was impossible to construct a CSS code if one of classical codes was of type IEEE802.16e with rate 1/2 and 2/3B.
We would like to report the discussion on its impossibility in this paper.
This is the first paper to analyze the possibility of a CSS code construction by using two irregular LDPC codes which are practically useful.

\section{Preliminaries}
\subsection{CSS codes}
Denote the binary field by $\mathbb{F}_2$.
\textbf{CSS (Calderbank-Shor-Steane) codes} are quantum codes constructed by a pair of classical linear codes $C$ and $D$ which satisfy the following condition (T), called \textbf{twisted relation}:
\begin{itemize}\label{twisted}
\item[(T)] $D^{\perp} \subset C$,
\end{itemize}
where $D^{\perp}$ is the \textbf{dual code} of $D$.
The dual code $D^{\perp}$ of $D$ is defined by the following:
$$D^{\perp} := \{ d' | d \times d'^{T} = 0, \forall d \in D \},$$
where $d'^{T}$ is the transposed vector of $d'$.
A function $\langle, \rangle$, defined by $\langle x, y\rangle := x \times y^{T}$, gives an inner product over the binary field $\mathbb{F}_2$.
The dual code $D^{\perp}$ is regarded as an orthogonal space of $D$ via $\langle, \rangle$.
Note that $D^{\perp}$ is equal to a linear code generated by a parity-check matrix $H_D$ of $D$.
In other words, 
$$
D^{\perp} = \{ x H_D | x \in \mathbb{F}_2^{m} \},
$$
where $m$ is the column size of $H_D$.

Details of the construction of a ``quantum'' CSS code by such a pair of ``classical'' codes $C$ and $D$ is written in \cite{QCQI}.

\begin{rem}
In general, a pair of linear codes forming a CSS code is denoted by $C_{1}$ and $C_{2}^{\perp}$.
For simplicity, we use $C$ (resp. $D$) instead of $C_{1}$ (resp. $C_{2}^{\perp}$).
\end{rem}

\subsection{Quasi-Cyclic LDPC Codes}
A quasi-cyclic LDPC code with circulant matrices (\textbf{QC-LDPC code}) is defined by a binary matrix $H$ of size $m$-by-$n$, where $n$ is the length of the code and $m$ is the number of parity check bits in the code \cite{Fossorier}.
The number of systematic bits is $k = n - m$.
The matrix $H$ is defined as:
$$
H =
\left[
  \begin{array}{cccc}
    P_{0,0}    & P_{0,1}     & \dots   & P_{0, L-1}   \\
    P_{1,0}    & P_{1,1}     & \dots   & P_{1, L-1}   \\
    \vdots     & \vdots      & \ddots  & \vdots          \\
    P_{J-1, 0} & P_{J-1,1}   & \dots   & P_{J-1, L-1}   \\
  \end{array}
\right]
$$
where $P_{i,j}$ is one of a set of $z$-by-$z$ circulant permutation matrices or a $z$-by-$z$ zero matrix and $z$ is a positive integer.
The matrix $H$ is expanded from a binary \textbf{base matrix} $H_b$ of size $J$-by-$L$, where $m= z J$ and $n = z L$.
The base matrix is expanded by replacing each $1$ in the base matrix with a $z$-by-$z$ circulant permutation matrix, and each $0$ with a $z$-by-$z$ zero matrix.
The circulant permutations used are circular right shifts, and the set of permutation matrices contains the $z \times z$ identity matrix and circular right shifted versions of the identity matrix.
Because each permutation matrix is specified by a single circular right shift, the binary base matrix information and permutation replacement information can be combined into a single compact model matrix $\mathcal{H}$.
The \textbf{model matrix} $\mathcal{H}$ is the same size as the binary base matrix $H_b$, with each binary entry $(i,j)$ of the base matrix $H_b$ replaced to create the model matrix $\mathcal{H}$.
Each 0 in $H_b$ is replaced by a symbol $\infty$ to denote a $z \times z$ all-zero matrix, and each $1$ in $H_b$ is replaced by a circular shift size $p(i,j) \ge 0$.
The model matrix $\mathcal{H}$ can then be directly expanded to $H$.
Denote a circulant permutation matrix of single shift by $I(1)$, i.e.
$$ I(1) = 
\left(
  \begin{array}{ccccc}
       & 1   &    &    &    \\
       &    &  1  &    &    \\
       &    &    & \ddots   &    \\
       &    &    &    &  1  \\
    1   &    &    &    &    \\
  \end{array}
\right),$$
and $I(b) = I(1)^{b}$.
A circulant permutation $P_{i,j}$ in $H$ is $I(p(i,j))$.

\begin{example}
Let $H$ be a parity-check matrix of size $6$-by-$10$ with the following form:
$$
H = 
\left(
  \begin{array}{cccccccccc}
 0 & 0 & 1 & 0 & 0 & 1 & 1 & 0 & 0 & 0 \\
 0 & 0 & 0 & 1 & 1 & 0 & 0 & 1 & 0 & 0 \\
 1 & 0 & 1 & 0 & 1 & 0 & 1 & 0 & 1 & 0 \\
 0 & 1 & 0 & 1 & 0 & 1 & 0 & 1 & 0 & 1 \\
 0 & 1 & 1 & 0 & 0 & 1 & 0 & 0 & 1 & 0 \\
 1 & 0 & 0 & 1 & 1 & 0 & 0 & 0 & 0 & 1 \\
  \end{array}
\right).$$
Then $H$ defines a quasi-cyclic LDPC code with circulant matrices of size $2$.
The base matrix $H_b$ associated to $H$ is a matrix of size $3$-by-$5$:
$$
H_b = 
\left(
  \begin{array}{ccccc}
 0 & 1 & 1 & 1 & 0 \\
 1 & 1 & 1 & 1 & 1 \\
 1 & 1 & 1 & 0 & 1 \\
  \end{array}
\right).$$
The model matrix $\mathcal{H}$ associated to $H$ is a matrix of size $3$-by-$5$:
$$
\mathcal{H} = 
\left(
  \begin{array}{ccccc}
 \infty & 0 & 1 & 0 & \infty \\
 0 & 0 & 0 & 0 & 0 \\
 1 & 0 & 1 & \infty & 0 \\
  \end{array}
\right).$$
\end{example}

Denote a set $\{0, 1, \dots, z-1\} \cup \{\infty\}$ by $[z]_{\infty}$.
A model matrix is a matrix over $[z]_{\infty}$.
We define a operator $-$ on $[z]_{\infty}$.
For $a, b \in \{0,1, \dots, z-1 \}$, we define $a - b$ as an usual integer operation modulo $z$.
For $a \in \{0, 1, \dots, z-1 \}$, we define $a- \infty = \infty - a = \infty - \infty = \infty$.
For vectors $v= (v_{0}, v_{1}, \dots, v_{L-1})$ and $u = (u_{0}, u_{1}, \dots, u_{L-1})$, we define $v - u = (v_{0} - u_{0}, v_{1} - u_{1} , \dots, v_{L-1} - u_{L-1})$.
For example, we have $(\infty, 0,1, 0, \infty) - (1, 0, 1, \infty, 0) = (\infty, 0, 0, \infty, \infty)$.

\subsection{Quasi-Cyclic LDPC codes of type IEEE802.16e}
In the standardization IEEE802.16e, quasi-cyclic LDPC codes are chosen as one of optimal error-correcting codes and six model matrices are written to define quasi-cyclic LDPC codes.
For each model matrix, 19 kinds of the size of cireculent matrices are chosen.
Totaly, 114 quasi-cyclic LDPC codes are chosen in IEEE802.16e standards.
These six model matrices are designed to satisfy the following conditions.
In this paper, we call a quasi-cyclic LDPC code (a LDPC matrix, a base matrix, and a model matrix) which satisfies the following condition \textbf{of type IEEE802.16e}.
\begin{rem}
The conditions below is not a characterization but a generalization of the 6 model matrices in IEEE802.16e.
In fact, the set of model matrices which satisfy the condition below contains the 6 model matrices in IEEE.
\end{rem}

The base matrix $H_b$ of a quasi-cyclic LDPC matrix is partitioned into two sections, where $H_{b1}$ corresponds to the systematic bits and $H_{b2}$ corresponds to the parity-check bits, such that $H_b = [ (H_{b1})_{J \times (L-J)} | (H_{b2})_{J \times J} ]$.

Section $H_{b2}$ is further partitioned into two sections, where $h_b$ is a vector whose weight is three, and $H'_{b2}$ has a dual-diagonal structure with matrix elements at row $i$, column $j$ equal to $1$ for $i=j$, $1$ for $i=j+1$, and $0$ elsewhere:
$$ H_{b2} = [ h_b | H'_{b2} ] =
\left[
  \begin{array}{c|cccccc}
    h_{b}(0)    & 1  &    &         &   &   \\
    h_{b}(1)    & 1  & 1  &         &   &   \\
    \vdots      &    & 1  & \ddots  &   &   \\
    \vdots      &    &    & \ddots  & 1 &   \\
    \vdots      &    &    &         & 1 & 1 \\
    h_{b}(J-1) &    &    &         &   & 1 \\
  \end{array}
\right].
$$
The base matrix has $h_{b}(0)=1, h_{b}(J-1)=1$, and a third value $h_{b}(j)=1$ with some $0 < j < J-1$.
In particular, the non-zero sub-matrices are circularly right shifted by a particular circular shift value. Each $1$ in $H'_{b2}$ is assigned a shift size of $0$, and is replaced by a $z \times z $ identity matrix when expanding to $H$.
The two located at the top and the bottom of $h_b$ are assigned equal shift sizes, and the third $1$ in the middle of $h_b$ is given an unpaired shift size.

\begin{example}
We pick up two base matrices $H_{b(1/2)}$ and $H_{b(2/3 B)}$ written  in the standardization IEEE802.16e.
The matrix $H_{b(1/2)}$ defines quasi-cyclic LDPC codes of rate $1/2$ and has the following form:
$$ H_{b(1/2)} = 
\left[
  \begin{array}{c}
	011000001100110000000000	\\
	010001110001011000000000	\\
	000111010001001100000000	\\
	101000001100000110000000	\\
	001000100110000011000000	\\
	000011010001100001100000	\\
	001100000110000000110000	\\
	011000100100000000011000	\\
	100011010001000000001100	\\
	000001010011000000000110	\\
	001100001100000000000011	\\
	100001010001100000000001	\\
  \end{array}
\right].
$$
The matrix $H_{b(2/3B)}$ defines quasi-cyclic LDPC codes of rate $2/3$ and has the following form:
$$ H_{b(2/3B)} = 
\left[
  \begin{array}{c}
	101010101010101011000000	\\
	010101010101010101100000	\\
	101010101010101000110000	\\
	010101010101010100011000	\\
	101010101010101000001100	\\
	010101010101010100000110	\\
	101010101010101010000011	\\
	010101010101010110000001	\\
  \end{array}
\right].
$$
\end{example}

\subsection{Preliminaries of the twisted condition for Quasi-Cyclic LDPC codes}
Recall that a CSS code is defined by two classical linear codes.
Denote the two classical codes by $C$ and $D$.
We denote a quasi-cyclic LDPC matrix, the model matrix, and the base matrix associated with $C$ by $H_C$, $H_{b(C)}$, and $\mathcal{H}_C$, respectively.
Similarly, we use the notation $H_D$, $H_{b(D)}$, and $\mathcal{H}_D$ to denote a quasi-cyclic LDPC matrix, the model matrix, and the base matrix associated with $D$, respectively.

In \cite{hagiwara}, a necessary and sufficient condition for QC-LDPC codes $C$ and $D$ to satisfy (T) in terms of the model matrices has been obtained.
We quote the necessary and sufficient condition from \cite{hagiwara}.
Denote $j$ th rows of the model matrices $\mathcal{C}$ and $\mathcal{D}$ by $c_{j}$ and $d_{j}$ , respectively.
We call a vector $v=(v_{0}, v_{1}, \dots, v_{L-1})$ over $\{0, 1, \dots, z-1\} \cup \{ \infty \}$ \textbf{multiplicity-even} if each symbol except for $\infty$ appears even times in $\{v_{0}, v_{1}, \dots, v_{L-1}\}$.
For example, a vector $(0,0,1,1,1,\infty,1)$ is multiplicity-even.

\begin{thm}[\cite{hagiwara} Prop. 3.1. and Sec. IV(D)]\label{iff:twisted}
Let $C$ and $D$ be quasi-cyclic LDPC codes.
The codes $C$ and $D$ satisfy (T) if and only if
$c_{j} - d_{k}$ is multiplicity even for any row $c_j$ of the model matrix $\mathcal{H}_C$ and any row $d_k$ of the model matrix $\mathcal{H}_D$.
\end{thm}

\section{A Necessary Condition for the Twisted Condition in Terms of the Base Matrices}
\begin{prop}\label{prop:smaller}
Let $C$ and $D$ be quasi-cyclic LDPC codes with circulant matrices of size $z$.
Let $y$ be a positive integer such that $y$ divides $z$.
Let $C'$ (resp. $D'$) be a quasi-cyclic LDPC code with the model matrix same as $C$ (resp. $D$) and with circulant matrices of size $y$.
If $C$ and $D$ satisfy the twisted condition then $C'$ and $D'$ satisfy the twisted condition.
\end{prop}
\begin{proof}
For any positive integers $a, b$, if $a = b \pmod z$ then $a \mod b \pmod y$.
It implies that if a vector $v$ over $[z]_{\infty}$ is multiplicity-even then the vector $v$ is multiplicity-even as a vector over $[y]_{\infty}$.

Let $c_i$ (resp. $d_i$) be the $i$th row of the model matrix $\mathcal{H}_C$ (resp. $\mathcal{H}_D$) of $C$ (resp. $D$).
By Theorem \ref{iff:twisted}, $c_j - d_k$ is multiplicity-even as a vector over $[z]_{\infty}$.
Thus $c_j - d_k$ is multiplicity-even as a vector over $[y]_{\infty}$.
\end{proof}

\begin{cor}
Let $C$ and $D$ be quasi-cyclic LDPC codes with circulant matrices.
Let $H_{b(C)}$ (resp. $H_{b(D)}$) be the base matrix of $C$ (resp. $D$).
Let $C'$ (resp. $D'$) be a linear code with a parity-check matrix $H_{b(C)}$ (resp. $H_{b(D)}$).
If $C$ and $D$ satisfy the twisted condition, then $C'$ and $D'$ satisfy the twisted condition, i.e. we have
$$ H_{b(C)} \times H_{b(D)}^{T} = 0.$$
\end{cor}
\begin{proof}
The base matrix $H_{b(C)}$ is a quasi-cyclic LDPC matrix which has the model matrix same as that of $C$ with circulant matrices of size $1$.
By Proposition \ref{prop:smaller}, $C'$ and $D'$ satisfy the twisted condition.

Remember that two linear codes satisfy the twisted condition if and only if these parity-check matrices $H_{1}$ and $H_{2}$ are orthogonal to each other i.e.
$$H_{1} \times H_{2}^{T} = 0. $$
Thus we have $ H_{b(C)} \times H_{b(D)}^{T} = 0.$
\end{proof}
\section{Non-Existance Results for a pair of QC-LDPC Codes of Type IEEE802.16e with Rate 1/2 and 2/3B}
The column (resp. row) weight distribution is defined as the distribution of Hamming weights of the columns (resp. the rows) for a given parity-check matrix.
The weight distribution is one of the important parameters for the design of LDPC codes.
In fact, the performance of LDPC codes with well-optimized weight distribution is very close to the asymptotic theoretical bounds \cite{richardson}.
Note that the weight distribution of the parity-check matrix of quasi-cyclic LDPC code with circulant matrices is the same as that of its base matrix.
There are $i_w$ columns (resp. rows) of Hamming weight $w$ in the base matrix if and only if there $i_w z$ columns (resp. rows) of Hamming weight $w$ in associated parity-check matrix, where $z$ is the size of the circulant matrices.

In this section, we discuss our problem, which is a construction of a pair of irregular LDPC codes to derive a CSS code, under the following conditions:
\begin{itemize}
\item Classical code $C$ and $D$ are quasi-cyclic LDPC codes with circulant matrices of size $z$.
\item The associated low-density parity-check matrices of $C$ and $D$ have the same ``row-weight'' distributions.
\item The weight of columns is more than or equal to 2. This condition arises from a standard decoding method, which is called a \textbf{sum-product decoding}. If one of a column-weight is less than 2, then the sum-product decoding does not work well \cite{gallager}.
\end{itemize}
In this paper, denote the condition above by (I).
The condition (I) does not impose the limitation on the base matrices.

In subsection \S \ref{ss:2/3B}, we discuss the impossibility on construction of $D$ under $H_{b(C)}:=H_{b(2/3B)}$.
Similarly, in subsection \S \ref{ss:1/2}, the impossibility on construction under $H_{b(C)}:=H_{b(1/2)}$ is discussed.

\subsection{Construction Impossibility on $D$ under $H_{b(C)}:=H_{b(2/3B)}$}\label{ss:2/3B}
Let $C$ be a quasi-cyclic LDPC code with the base matrix $H_{b(C)} := H_{b(2/3B)}$.

It is required that the base matrix $H_{b(D)}$ of $D$ satisfies:
$$ H_{b(C)} \times H_{b(D)}^{T} = 0.$$
It implies that any row $x \in \mathbb{F}_2^{24}$ of $H_{b(D)}$ satisfies:
$$H_{b(C)} x = 0.$$
Define a matrix $M$ as follows:
$$
M := 
\left[
  \begin{array}{c}
	11111111	\\
	01111111	\\
	00111111	\\
	00011111	\\
	00001111	\\
	00000111	\\
	00000011	\\
	11111110	\\
  \end{array}
\right].
$$
Then, for a vector $x \in \mathbb{F}_{2}^{24}$, we have $$H_{b(C)} x = 0 \iff M H_{b(C)} x = 0.$$
By an easy calculation, we obtain
$$
M H_{b(C)} =
\left[
  \begin{array}{c}
	000000000000000010000000	\\
	101010101010101001000000	\\
	111111111111111100100000	\\
	010101010101010100010000	\\
	000000000000000000001000	\\
	101010101010101000000100	\\
	111111111111111100000010	\\
	010101010101010100000001	\\
  \end{array}
\right].
$$
By considering the first and the fifth rows, we obtain a necessary condition $x_{17}=x_{21}=0$ for $H_{b(C)} x=0$, where $x=(x_{1}, x_{2}, \dots, x_{24})$.

Thus the weight of 17th column and 21st column of $H_{b(D)}$ must be 0, in particular is less than 2.
Therefore it is impossible to construct $D$ which satisfies the condition (I) under $H_{b(C)} = H_{b(2/3 B)}$.
By the result, we immediately have the 
\begin{cor}
Let $C$ be one of 19 quasi-cyclic LDPC codes which has the ``base matrix'' written as rate 1/2 codes in IEEE802.16e.
Then there is no quasi-cyclic LDPC code $D$ which satisfies (I).
\end{cor}
\begin{proof}
$C$ is a particular example which satisfies our condition.
\end{proof}
\subsection{Construction Impossibility on $D$ under $H_{b(C)}:=H_{b(1/2)}$}\label{ss:1/2}
Let $C$ be a quasi-cyclic LDPC code with its base matrix $H_{b(C)} := H_{b(1/2)}$.

By the twisted condition, for the base matrix $H_{b(D)}$ of a quasi-cyclic LDPC code $D$, the following equation is required:
$$ H_{b(C)} \times H_{b(D)}^{T} = 0.$$

It implies that any row $x \in \mathbb{F}_2^{24}$ of $H_{b(D)}$ satisfies:
$$H_{b(C)} x = 0.$$
The weight of each row of $H_{b(D)}$ is 6 or 7.
By the definition of a quasi-cyclic LDPC code of type IEEE802.16e, two non-zero entries of any row of $H_{b(D)}$, except for the below, appear in adjacent position to each other.
The weight of the right-side 11 bits of a row of $H_{b(D)}$ is one or two and we denote the weight of the right-side by $\mathrm{wt|_{13-}}$.

Define three sets $X_1, X_2$ and $X_3$:
$$ X_1 := \{ x \in \mathbb{F}_2^{24} | \mathrm{wt}(x) = 6 \text{ or } 7 \},$$
$$ X_2 := \{ x \in \mathbb{F}_2^{24} | x_{i}=x_{i+1}=1 \text{ for some 14 } \le i \le 23\},$$
$$ X_3 := \{ x \in \mathbb{F}_2^{24} | \mathrm{wt|_{13-}}(x) = 1 \text{ or } 2\}.$$
For simplicity, we denote $X := X_1 \cap X_2 \cap X_3$.
Then each row of $H_{b(D)}$ belongs to the set $X$.

We can determine the set $X$ by using a personal computer.
In fact, we verify $X$ consists of six elements.
The right-side 12-bits of each element of $X$ have the form $(0, 0, 0, 0, 0, 0, 0, 0, 1, 1, 0, 0)$.
On the other hand, eleven kinds of the right-side 12-bits are required to satisfy (I).
It shows that it is impossible to construct $H_{b(D)}$ which satisfies (I) under the assumption $H_{b(C)}:=H_{b(1/2)}$.

By the result, we immediately have the 
\begin{cor}
Let $C$ be one of 19 quasi-cyclic LDPC codes which has the ``base matrix'' written as rate 2/3B codes in IEEE802.16e.
Then there is no quasi-cyclic LDPC code $D$ which satisfies (I).
\end{cor}
\begin{proof}
$C$ is a particular example which satisfies our condition.
\end{proof}

\section{Conclusion}
In this paper, we discuss the conditions to construct a pair of quasi-cyclic LDPC codes, in particular of type IEEE802.16e, to derive a CSS code.
The key of our discussion is to analyze the base matrices of quasi-cyclic LDPC codes.
By our research, we find the impossibility to construct a pair of quasi-cyclic LDPC codes to derive a CSS code under conditions that (I) holds and one of the base matrices is the same as an IEEE802.16e LDPC base matrix with rate 1/2 or 2/3B.
Note that our results are more general.
The 32 quasi-cyclic LDPC codes, which have the model matrice written as rate 1/2 and rate 2/3B in IEEE802.16e, are typical examples of our results.

We should remember that six base matrices, totally 96 quasi-cyclic LDPC codes, are discribed in the standardization form of IEEE802.16e.
Hope has been left for the other four LDPC codes, totally 76 quasi-cyclic LDPC codes.
We leave these posibilities as open problems.

\section*{Acknowledgement}
This research was partially supported by Grants-in-Aid for Young Scientists (B), 18700017, 2006.



%

\end{document}